# Compound Attrition Games: A Unified Model for Inter- and Intra-Coalition Rivalry

**Authors**: Madjid Eshaghi Gordji*, Mohamad Ali Berahman

**Affiliations:**
Department of Mathematics Education, Faculty of Mathematics, Statistics and Computer Science, University of Semnan, Semnan, Iran

Corresponding author: meshaghi@semnan.ac.ir

## Abstract

Strategic competitions in the real world—from standard wars to geopolitical rivalries—often involve groups of actors forming coalitions to compete against rival groups. These contests are not simple battles between unified entities; rather, they are "wars within wars," where coalitions fight each other while simultaneously grappling with internal conflicts over resources and strategy. However, existing game-theoretic models typically treat these two layers—inter-coalition rivalry and intra-coalition competition—separately. This paper addresses this fundamental gap by introducing **the Compound Coalition–Attrition Game** (**CCAG**), a unified theoretical framework that formally integrates a war of attrition between coalitions with a simultaneous war of attrition within each coalition. In our model, the collective endurance of a coalition in external competition is endogenously determined by the strategic choices of its members, who internally compete for shares of the eventual spoils. We prove the nonexistence of pure-strategy equilibria and characterize the unique mixed-strategy Nash equilibrium for both layers of the game. Our analysis reveals powerful feedback loops: the prospect of external victory intensifies internal competition, while internal discord weakens the coalition's external standing—producing virtuous cycles of success or vicious cycles of fragmentation. We validate the model through a detailed case study comparing the traditional market (including gold, copper, and silver) with the cryptocurrency market (including Bitcoin, Ethereum, and Solana), using real data from 2018 to 2023 in a simulation framework. The results demonstrate its broad applicability in industrial strategy, corporate decision-making, and geopolitical competition. The Compound Coalition–Attrition Game (CCAG) thus offers a powerful new tool for understanding and predicting outcomes in complex, multilayered strategic environments.

## Introduction

The study of strategic interactions, known as game theory, provides a mathematical framework for analyzing situations in which the outcome for each player depends on the decisions made by others. The theoretical foundation of this field was established by the classic work Theory of Games and Economic Behavior authored by **von Neumann and Morgenstern (1944)**[1]. This seminal work not only created a common language across economics, political science, and biology but also fundamentally transformed the analytical foundations of rational decision-making.
Subsequently, **Nash (1951)** [2]advanced equilibrium analysis for multiplayer games by introducing the concept of Nash equilibrium. Defined as a state in which no player can increase their payoff through a unilateral change in strategy, Nash equilibrium became central to numerous strategic models. This framework successfully explained phenomena ranging from oligopolistic competition to nuclear deterrence.
However, many real-world scenarios also involve cooperation. This led to the development of the branch of coalition games, pioneered by researchers such as **Shapley (1953)**[3] and **Scarf (1967)**[4].
Cooperative games focus on the question of how groups of players can collaborate to achieve greater benefits and how these benefits should be fairly allocated. The Shapley value (**Shapley, 1953**)[3] is one of the most well-known solution concepts for distributing cooperative gains, based on principles of fairness, symmetry, and efficiency.
The concept of the core, introduced by **Scarf (1967)**[4], characterizes the set of allocations that cannot be improved upon by any coalition. **Debreu and Scarf (1963)**[5] established conditions for the non-emptiness of the core in public economies.
**Shenoy (1979)**[6] developed abstract models and algebraic-geometric tools for predicting coalition formation. **Ray (2007)**[7] proposed an integrative perspective bridging cooperative and non-cooperative games, which analyzes the actual process of coalition formation.
Recent studies have further advanced the analysis of coalition formation. **Montero (2023)**[8] examined coalition formation in the presence of externalities, demonstrating that equilibrium paths can be diverse even under conditions of initial symmetry. **Bachrah et al.(2011)**[9] investigated optimal coalition structures within graphical games**. Hjørungnes (2009)**[10] explored applications of cooperative game theory in communication networks.
From a dynamic perspective, **Ali and Liu (2019)**[11] proposed a framework for coalition stability in repeated games using promises and punishments. **Bolton and Brosig-Koch (2012)**[12], through experimental studies, showed that communication among players can enhance coalition efficiency. **Okada and Riedl (2005)**[13] analyzed the impact of mutual preferences on coalition stability and efficiency. **Calvo and Gutiérrez (2010)**[14] introduced a correlation-based allocation rule that promotes both stability and fairness.
Other studies, such as **Sugiyama et al. (2021)**[15], have examined coalition formation in hunter-gatherer societies from an evolutionary perspective. **Magaña and Carreras (2018)**[16] proposed a combination of the Shapley value with the strong Nash equilibrium to analyze coalition stability.
**The War of Attrition** was first introduced by **John Maynard Smith (1974)**[17] within the framework of evolutionary biology to explain competitive behaviors of animals over scarce resources. In this model, each player decides how long to remain in the competition; staying longer increases the probability of winning but also accumulates costs.
**Bishop and Cannings (1978)**[18] extended this model to incorporate diverse reward and cost functions as well as multiplayer games. **Tullock (1967)**[19], by introducing the concept of rent competition, established the theoretical connection between the War of Attrition and economic contest models.
**Skaperdas (1996)**[20] formalized the framework of Contest Success Functions (CSF).
**Baye, Kovenock, and de Vries (1996)**[21] provided a detailed analysis of all-pay auctions. **Kreps and Wilson (1982)**[22] examined the concept of reputation in games with incomplete information**. Isaacs (1999)**[23] introduced differential game tools that are applicable to continuous modeling of the War of Attrition.
**Hammerstein and Parker (1982)**[24] investigated the impact of asymmetric abilities on equilibrium outcomes. **Haigh and Rose (1980)**[25] explored the connection between the War of Attrition and evolutionary auctions. **Nalebuff and Riley (1985)**[26] analyzed asymmetric equilibria under private information. **Leininger (1989)**[27] modeled the processes of escalation and cooperation over time.
**Konrad and Kovenock (2009)**[28] examined multi-battle wars of attrition. **Décamps et al. (2022)**[29] analyzed Markovian mixed strategies with asymmetric information and stochastic payoffs. **Gieczewski (2025)**[30] developed an evolutionary model incorporating state-dependent costs. **Li and Zhang (2023)**[31] studied multi-player wars of attrition with incomplete and asymmetric information. **Myatt (2024)**[32] investigated the effect of perceived power. **Georgiadis et al. (2022)**[33] identified conditions under which wars of attrition do not occur.
espite these advancements, a significant gap remains: the absence of a model that simultaneously captures coalition formation, intra-coalition competition, and inter-coalition competition within a unified and dynamic framework. Cooperative games tend to simplify the internal dynamics following coalition formation, while wars of attrition treat each side as a cohesive unit.
In real-world scenarios, however, these two dimensions are deeply interdependent: success in external competition can alter incentives for internal cooperation, while internal conflicts may undermine the chances of external success.
The compound coalition-attrition games (CCAG) address this gap by incorporating two layers:

- Inter-coalition layer: War of attrition competition between coalitions.
- Intra-coalition layer: War of attrition competition among members within a coalition, which determines its overall endurance.

Theoretical analysis of the model demonstrates that mixed-strategy equilibria at both layers mutually influence each other, and feedback loops can generate reinforcing cycles of success or failure.
To test the model, a simulation was conducted involving competition between the traditional market (gold, oil, copper) and the cryptocurrency market (Bitcoin, Ethereum, Solana). In the first layer, these two markets engaged in inter-coalition war of attrition competition. In the second layer, the three cryptocurrencies competed for the final investment choice.
Data from 2018 to 2023 demonstrated that the model can accurately reproduce the competition trajectory and predict conditions under which the winner changes. A counterfactual analysis was also performed to examine the effects of variations in resources or costs on the outcomes.

**Theoretical Foundations: Wars of Attrition and Cooperative Games**

This section is divided into two subsections, each drawing on key academic contributions that have shaped the formal modeling of strategic interaction and coalition formation.Definition and Theoretical Basis The War of Attrition game was first introduced by **John Maynard Smith (1974)**[17] within the context of evolutionary biology, aiming to model animal behavior in conflicts over limited resources. In this framework, players must decide how long they are willing to persist in a contest. The longer one remains in the game, the higher the chance of winning, but the incurred cost also increases with time.Consider two players, $i$ and j, competing over a resource of value V. Each chooses a time $t_i$ and $t_j$ to remain in the contest. The rules are as follows:

- Cost of staying in the game:
  $C(t_i) = t_i$ (i.e., the cost is linearly proportional to time).
- Outcomes:
  if $t_i > t_j$, then player $i$ wins and receives net payoff $V - t_i$, while player j loses and incurs a cost of $-t_j$ .
- If $t_i = t_j$, the prize is split equally between the two players.

In the classic Nash equilibrium of the two-player game, each player randomizes their quitting time using an exponential distribution. The equilibrium cumulative distribution function (CDF) for the withdrawal time is given by:

$$F(t) = 1 - e^{-\frac{t}{V}}$$

(This expression applies to the two-player case. For $n$ players, the distribution changes accordingly.)
Cooperative games constitute a branch of game theory in which players are allowed to form coalitions and coordinate their strategies to achieve collectively beneficial outcomes. This paradigm was initially developed by von Neumann and Morgenstern (1944) in their foundational work on game theory.
Formal Structure:

- Set of players:
  $N = \{1, 2, \ldots, n\}$
- Characteristic function:
  $v : 2^N \to \mathbb{R}$

assigns a value to each coalition $S \subseteq N$ , representing the total payoff that the coalition can secure through cooperation
A central method for fairly distributing the total payoff among players based on their marginal contributions to all possible coalitions. Defined by **Lloyd Shapley (1953)[3]**, the Shapley value for player $i$ is calculated as:

$$\phi_i(v) = \sum_{S \subseteq N \setminus \{i\}} \frac{|S|!.(n-|S|-1)!}{n!} \left[ v(S \cup \{i\}) - v(S) \right]$$

The *core* is the set of payoff allocations in which no subgroup of players has an incentive to break away and form a separate coalition. A vector $x \in \mathbb{R}^n$ lies in the core if:

$$\sum_{i \in N} x_i = v(N) \quad , \quad \textit{and} \quad \sum_{i \in S} x_i \geq v(S) \quad \forall S \subseteq N$$

In contrast to non-cooperative games, which focus on individual strategy optimization and equilibrium, cooperative games emphasize negotiation, coalition stability, and collective resource allocation. Stability and fairness become central concerns in determining acceptable outcomes for all involved parties.

## Main Result

In this model, we first present the general framework of the game, followed by the definitions of the players and coalitions involved. The player population is denoted by

$$N = \{1, 2, \ldots, n\}$$

representing a set of investors who are eventually grouped into coalitions. These coalitions represent organized groups that jointly engage in competition to attract investment and accumulate economic power.

A coalition is defined as a subset of players $S_k \subseteq N$, acting as a unified organizational or economic entity within the game.

Coalitions are formed such that they are mutually exclusive:

$$S_i \bigcap Sj = \varnothing \quad s.t \quad i \neq j$$

and collectively exhaustive, i.e., all players are assigned to some coalition:

$$\bigcup_{k=1}^{m} S_k = N$$

where *m* denotes the total number of coalitions.

The proposed game exhibits a two-layered competitive structure that unfolds simultaneously:

**Layer 1: Inter-Coalition Competition**

At this level, coalitions act as unified strategic entities competing for investment, market share, or economic dominance. This competition is modeled as a *war of attrition* among coalitions, where the objective is to secure a major prize—such as market leadership or control over critical resources.

**Layer 2: Intra-Coalition Competition**

Within each coalition, individual members engage in a secondary competition for a larger share of the coalition's internal resources. Players strive to maximize their individual payoffs by balancing two objectives: contributing to the coalition's success in the external competition, and securing a favorable allocation of the rewards internally. This intra-group dynamic reflects tensions over effort, resilience, and the distribution of collective gains.

Resilience, denoted by $t_i \in [0, \infty)$, is a strategic variable determined by each player $i$ within their respective coalition. It captures the player's level of effort, time investment, or capacity to endure economic pressure in order to enhance the coalition's chances of success and secure a share of the available resources.

A higher value of $t_i$ increases the player's influence and share in the intra-coalition competition. However, exerting resilience incurs costs—including opportunity costs, financial risks, and psychological burdens—which must be incorporated into the player's utility function.

**Player Strategies**

Each player operates at two levels of decision-making:

- Coalition Selection $S_i$:

  A discrete decision to join one of the available coalitions, where $S_i \in \{S_1, S_2, \ldots, S_m\}$.

- Resilience Selection $t_i$:

  A continuous decision regarding the level of effort and participation within the chosen coalition, represented by a non-negative real number $t_i \in [0, \infty)$.

At the initial stage, each player $i \in N$ must choose one of the available coalitions to join. This selection represents a discrete decision based on the player's rational assessment of expected outcomes. It is assumed that each player chooses a coalition that maximizes their expected utility.
In this model, it is assumed that players make informed decisions by assessing various characteristics of each coalition, including its number of members, competitive strength, likelihood of winning the inter-coalition contest, and the expected intra-coalition payoff.
When choosing among potential coalitions, player *i* considers several critical factors:

- Probability of Success in the Inter-Coalition Contest**:**
  Coalitions with a higher likelihood of winning the external competition are generally more attractive to players seeking greater expected returns.
- Intra-Coalition Structure and Reward Distribution:
  Players assess whether the internal dynamics of a coalition allow for a favorable share of the collective reward. Factors such as internal competition intensity and the fairness of payoff allocation play key roles in this evaluation.
- Intra-Coalition Effort Costs:
  The level of resilience or effort required to secure a meaningful share within the coalition is also crucial. High internal competition may demand significantly greater effort, which could offset potential gains.

We assume that player *i* evaluates their expected utility from membership in coalition $S_k$ as follows:

$E\left[\pi_i\left(S_k\right)\right] =$ Probability of $S_k$ winning × Player *i*'s share of the reward−Cost of resilience.

In other words, if a coalition has a high probability of winning and moderate internal competition (allowing the player to secure a reasonable share of resources), the player is inclined to join it.
Conversely, if the intra-coalition competition is intense, the associated costs may outweigh the benefits, motivating the player to prefer an alternative coalition.
Subsequently, each player *i* must decide which of the available coalitions $S_k$ to join. It is assumed that each player acts rationally, aiming to maximize their utility (net payoff). Accordingly, the player evaluates their expected utility within each coalition $S_k$ and subsequently selects the coalition that offers the highest expected payoff.
Thus, the player's decision-making process is based on comparing the utility functions corresponding to all potential coalitions. This utility function is formally defined in the following section.
Each player *i* initially faces a set of possible coalitions:

$$S_i \in \{S_1, S_2, \ldots, S_m\}.$$

The player's objective is to select the coalition that maximizes their net payoff (utility). Accordingly, the player's decision rule is defined as:

$$S_i^* = \arg\max_{S_k \in S} \pi_i(S_k)$$

**Definition of Parameters:**

| ***Parameters*** | ***Definition*** |
|---|---|
| $S_i^*$ | Optimal coalition chosen by player *i* |
| $S$ | The set of all possible coalitions |
| $S_k$ | A particular coalition within the game |
| $\pi_i(S_k)$ | Expected payoff of player *i* when joining coalition $S_k$ |

**Table1** . lists the main symbols and their definitions used in modeling players' coalition selection decisions.

**Utility Function of Player *i* in Coalition $S_k$**

When player *i* decides to join coalition $S_k$, their utility from participating in that coalition is given by:

$$\pi_i\left(S_k\right)=\left(\frac{t_i\ a_i}{\sum_{j\in S_k} t_j a_j}\right).R_k-C(t_i) \quad (1)$$

In this expression, $\pi_i(S_k)$ denotes the net payoff of player *i* upon membership in coalition $S_k$. The term $t_i$ represents the resilience level of player $i$, reflecting their capacity for effort or endurance under pressure, while $a_i$ measures their individual effectiveness within the coalition. The denominator $\sum_{j\in S_k} t_j a_j$ captures the aggregate effective effort exerted by all members of the coalition. $R_k$ corresponds to the total reward available to coalition $S_k$ upon winning against rival coalitions. Finally, $C(t_i)$ is the cost function associated with player *i's* level of resilience, which is assumed to be strictly increasing.

The share of player *i* from the total reward $R_k$ is calculated as the ratio of the player's individual effective effort to the total effective effort of the coalition members:

$$\left(\frac{t_i\ a_i}{\sum_{j\in S_k} t_j a_j}\right)$$

The cost function $C(t_i)$ is typically assumed to be an increasing function, such as $C(t_i)=\lambda t_i^2$ or similar forms, reflecting that higher levels of effort incur greater costs for the individual. Consequently, the player balances the relative gain from participation against the associated cost and makes decisions accordingly.

Once each player selects a coalition based on the maximization of their utility function (Equation 1), the formed coalitions enter into a competitive interaction. In this contest, only one coalition emerges victorious, earning the total reward $R_k$, while all rival coalitions receive no payoff.

To model the probability of a coalition $S_k$ winning, we define a collective power function. The power of a coalition is quantified as the sum of the effective efforts of its members:

$$Power\left(S_k\right)=\sum_{j\in S_k} t_j a_j$$

The probability that coalition $S_k$ wins the competition is then determined as the ratio of its power to the total power of all competing coalitions:

$$P_k^{win}=\frac{\sum_{j\in S_k} t_j a_j}{\sum_h \sum_{j\in S_h} t_j a_j}$$

Here, $S_h$ represents all coalitions in the system, and the denominator reflects the aggregated power of all participating coalitions.

For a player *i* who has joined coalition $S_k$, the actual payoff depends not only on their share within the coalition but also on the probability of that coalition's success. Hence, the expected utility of player *i* is given by:

$$E\left[\pi_i\left(S_k\right)\right]=P_k^{win}.\left(\frac{t_i\ a_i}{\sum_{j\in S_k} t_j a_j}.R_k\right)-C(t_i)$$

This formulation captures a more realistic representation of the decision-making process: although player *i* initially chooses their coalition under the assumption of victory (as per Equation 3.1), only a fraction of the anticipated reward is realized—proportional to the coalition's probability of winning $P_k^{win}$.

This final utility function plays a central role in equilibrium analysis, sensitivity exploration, and mechanism design within strategic coalition formation frameworks.

After player *i* selects coalition $S_k$ based on the maximization of their expected utility, they enter the second layer of decision-making—namely, the intra-coalition competition to join one of the existing sub-coalitions within $S_k$.

Each coalition $S_k$ is subdivided into multiple sub-coalitions (or specialized branches/market segments), denoted by $G_l \subseteq S_k$, each representing a distinct competitive track or a specific market. Player iii must then decide which sub-coalition $G_l \subseteq S_k$ to join in order to maximize their expected payoff. The utility function of player iii within sub-coalition $G_l$ is defined as follows:

$$\pi_i^{(G_l)} = P_{G_l}^{win} . \left( \frac{t_i a_i}{\sum_{j \in G_l} t_j a_j} . R_k^{(G_l)} \right) - C(t_i)$$

Where:

- $G_l \subseteq S_k$ : denotes a sub-coalition $G_l$ as a subset of the main coalition $S_k$.
- $\pi_i^{(G_l)}$ : represents the final utility of player *i* if they join sub-coalition $G_l$ .
- $R_k^{(G_l)}$ : refers to the reward allocated to sub-coalition $G_l$ from the total reward $R_k$ of the main coalition.
- $P_{G_l}^{win}$ : denotes the probability that sub-coalition $G_l$ will win the intra-coalition competition**.**

This formulation reflects the player's strategic evaluation of both inter- and intra-coalition dynamics in pursuit of maximal gain.

**Nash Equilibrium Analysis in the Two-Stage Model**

This section presents an analysis of the Nash equilibrium within the proposed two-stage model. The model comprises two levels of decision-making: first, players decide whether to join a particular coalition; second, within the selected coalition, they compete to secure a share of the collective reward. In this framework, a Nash equilibrium is defined as a set of strategies in which no player has an incentive to unilaterally deviate from their chosen course of action.

**Nash Equilibrium in the Coalition Formation Stage**

Let the set of players be denoted by $N = \{1, 2, \ldots, n\}$. Each player $i \in N$ must choose one of the two possible coalitions, denoted by $S_1$ and $S_2$ .

A coalition profile $S = (S_1 \ , S_2 \ )$ constitutes a Nash equilibrium if and only if, for every player $i \in N$ , the following condition holds:

$$\pi_i (S_k) \geq \pi_i (S_{k'}) \quad \forall k' \neq k$$

This means that no player can achieve a higher utility by unilaterally switching from their current coalition to the other one.

The existence of such an equilibrium necessitates consideration of the interdependence among players' decisions, since each individual's relative payoff is contingent upon the total collective effort. This interdependence may give rise to multiple equilibria, which can, in some cases, be unstable.

**Nash Equilibrium in the Intra-Coalition Competition Stage**

After the formation of coalitions, an attrition competition begins within each coalition. The intra-coalition game is a type of rent-seeking game in which each player selects an effort level $t_i$ to maximize their share of the collective reward.

We assume that the effort cost function for player iii is convex and increasing, defined as:

$$C(t_i) = c_i t_i^2, \quad c_i > 0$$

The payoff function for player $i \in S_k$ in this competition is given by:

$$\pi_i(t_i, t_{-i}) = \left( \frac{t_i a_i}{\sum_{j \in S_k} t_j a_j} .R_k \right) - c_i t_i^2$$

**Nash Equilibrium in Intra-Coalition Effort Games**

Consider the coalition $S_k$ where each player $i \in S_k$ chooses effort level $t_i \geq 0$ to maximize their utility (previously defined in Equation (2)).

A profile of efforts $t^* = \{t_i^*\}_{i \in S_k}$ is a **Nash equilibrium** if no player can increase their utility by unilaterally changing their effort, i.e.,

$$\pi_i\left(t_i^*, t_{-i}^*\right) \geq \pi_i\left(t_i, t_{-i}^*\right), \quad \forall t_i \geq 0, \ \ \forall i \in S_k$$

To characterize this equilibrium, we use the first-order condition:

$$\frac{\partial \pi_i}{\partial t_i}\left(t_i^*, t_{-i}^*\right) = 0 \ ,$$

which yields a system of nonlinear equations:

$$\frac{a_i R_k \sum_{j \neq i} t_j^* a_j}{\left(\sum_{j \in S_k} t_j^* a_j\right)^2} - 2 c_i t_i^* = 0 \ , \quad \forall i \in S_k$$

Due to strategic interactions and the nonlinear nature of these equations, explicit closed-form solutions are generally unavailable, necessitating numerical methods for equilibrium computation.

**Non-Existence of Nash Equilibrium in Pure Strategies**

In the intra-coalition game, each member $i \in S_k$ adjusts their behavior by selecting a resilience level $t_i \in [0,1]$ in order to maximize their individual utility. The utility function of player $i$ is defined as follows:

$$\pi_i = \frac{t_i a_i}{\sum_{j \in S_k} t_j a_j} - c\left(t_i\right) \qquad (2)$$

where $a_i > 0$ represents the relative importance of member $i$ within the coalition, and $c\left(t_i\right)$ is a continuously differentiable cost function associated with resilience. This function satisfies the following conditions: $c\left(0\right) = 0$ , $c'(t_i) > 0$ , and $c''(t_i) > 0$ .

We now demonstrate that a Nash equilibrium in pure strategies may not exist in this game. To do so, it suffices to examine conditions under which there exists no strategy vector $t = (t_1, \ldots, t_n) \in [0,1]^n$ such that no player has an incentive to deviate unilaterally.

Assume that all coalition members choose the same level of effort, i.e., $t_i = t > 0$ for all $i$. In this case, the utility of each player is given by:

$$\pi_i = \frac{a_i}{\sum_{j \in S_k} a_j} - c\left(t\right)$$

This utility is independent of $t_i$ , while the cost function $c(t)$ is strictly increasing; therefore, each player has an incentive to reduce their effort level $t_i$ to minimize their cost. If all players choose $t_i = 0$, the denominator in the utility function becomes zero, rendering the utility undefined. This demonstrates that no strategy profile exists in which no player has an incentive to deviate, and consequently, no pure-strategy Nash equilibrium exists.

This non-existence of equilibrium stems from the competitive, ratio-based structure of the utility function combined with the simultaneous incentive to minimize costs. Such a structure leads to internal instability in the allocation of efforts within the coalition, reflecting the inherent tensions of resource competition among group members.
Considering the non-existence of a pure-strategy Nash equilibrium in the intra-coalition attrition game, it becomes imperative to investigate the concept of mixed-strategy Nash equilibrium, which facilitates the existence of stable equilibria within this framework.

**Introduction and Definition of Mixed-Strategy Nash Equilibrium**

Given the absence of a pure-strategy Nash equilibrium in the intra-coalition attrition game, we turn our attention to the concept of mixed-strategy Nash equilibrium. A mixed-strategy equilibrium occurs when each player selects a probability distribution over the set of possible effort levels such that no player has an incentive to unilaterally deviate from their chosen distribution.
Formally, the mixed strategy for each player $i \in S_k$ is defined as a probability distribution vector over the effort space $T_i \subseteq \mathbb{R}^+$. The mixed-strategy profile

$$\sigma^* = \left(\sigma_i^*\right)_{i \in S_k}$$

constitutes a mixed-strategy Nash equilibrium if and only if:

$$\forall i \in S_k,\ \mathrm{E}_{\sigma_{-i}^*}\left[\pi_i\left(\sigma_i^*, \sigma_{-i}^*\right)\right] \geq \mathrm{E}_{\sigma_{-i}^*}\left[\pi_i\left(\sigma_i, \sigma_{-i}^*\right)\right], \quad \forall \sigma_i \in \Delta(T_i),$$

where $\Delta(T_i)$ denotes the space of probability distributions over $T$, and $\pi_i$ is the utility function of player $i$.

**Existence and Uniqueness of Mixed-Strategy Equilibrium**

According to the general existence theorem for mixed-strategy Nash equilibria **(Nash, 1950)**[2], in any game with a finite number of players and compact, convex strategy spaces—where utility functions are continuous and quasi-concave—at least one mixed-strategy equilibrium is guaranteed to exist.
In the context of the proposed model, each player's effort strategy space is continuous and convex, and the utility function is continuous and concave in strategic parameters. These conditions, supported by the theoretical results of continuous games, ensure the existence of a mixed-strategy equilibrium.
Regarding the uniqueness of the mixed-strategy equilibrium, due to the nonlinear structure and the interdependence of utility functions, uniqueness is generally not guaranteed and depends on numerical analysis and specific model assumptions. Nevertheless, the study of equilibrium stability can help distinguish more plausible and robust equilibria.
The existence of feedback loops between intra-coalition and inter-coalition contests gives rise to complex dynamic behaviors that influence the nature and stability of mixed-strategy equilibria. To assess the stability of these equilibria, we employ the framework of evolutionary game dynamics, such as replicator dynamics.
Dynamic modeling of these contest games reveals the following insights:

- Individual heterogeneity in resilience and effort cost leads to persistent fluctuations in players' choice of effort levels.
- Mixed-strategy equilibria may emerge as cyclical or quasi-stable, whereby strategic shifts among players induce long-term oscillations in coalition structure and resource allocation.
- These complex dynamics can result in the formation of positive feedback cycles of coalition success or, conversely, in coalition collapse over time.

Overall, a dynamic analysis of mixed-strategy equilibria provides a powerful framework for understanding long-term behavior and forecasting strategic outcomes in compound attrition games.

## Case Study and Simulation

To evaluate the practical applicability of the proposed theoretical framework, we conduct a case study on the strategic investment choice between two major financial markets: the cryptocurrency market and the traditional market. In this analysis, each market is modeled as a competing coalition at the inter-coalition level, while each coalition comprises a set of specific assets competing among themselves at the intra-coalition level. The game structure is formulated in accordance with the two-layer attrition model as follows:

**Stage One (Inter-Coalition Competition):**

The investor selects one of the two coalitions, the cryptocurrency market $S_{crypto}$ or the traditional market $S_{traditional}$. The total payoff of each coalition is calculated through an index based on the aggregate relative resilience of its members compared to those of the opposing coalition. This payoff is determined via a comparative function derived from the theoretical model and is normalized to unity for simplicity $(R_k = 1)$.

**Stage Two (Intra-Coalition Competition):**

Following the investor's coalition choice, the members within the selected coalition engage in an intra-coalition attrition game to compete for their share of the total payoff $R_k$. Each asset strategically decides its effort level $t_i$, which incurs a cost and is analyzed through the attrition game framework.

In this case study, the cryptocurrency market includes three major assets: Bitcoin, Ethereum, and Solana, while the traditional market comprises Gold, Copper, and Brent Crude Oil. The strategic effort level of each asset, modeled as a continuous non-negative variable $t_i \in \mathbb{R}_{\geq 0}$, is here estimated using the volatility-adjusted Sharpe ratio calculated over the historical period 2018–2023. This metric simultaneously weighs both the stability and relative returns of each asset, aligning well with our theoretical interpretation of resilience.

Subsequently, leveraging these data, the attrition model at both inter- and intra-coalition levels is simulated to numerically investigate the investor's strategic behavior and asset interactions within the proposed framework.

**Data Collection and Preparation**

To empirically evaluate the proposed composite attrition game framework, daily closing price data for selected assets in the cryptocurrency and traditional markets were retrieved from Yahoo Finance. Specifically, the cryptocurrency coalition comprises Bitcoin (BTC-USD), Ethereum (ETH-USD), and Solana (SOL-USD), while the traditional market coalition includes Gold (GC=F), Copper (HG=F), and Brent Crude Oil (BZ=F). The dataset covers the period from January 1, 2018, to December 31, 2023.

Data acquisition was conducted using the Python library yfinance, which facilitates access to financial data from Yahoo Finance. The downloaded data consist of daily closing prices organized into a Pandas DataFrame structure.

To estimate the strategic resilience of each asset, a volatility-adjusted Sharpe ratio was employed. This metric measures the risk-adjusted return by accounting for price volatility, providing a precise evaluation of the relative performance of the assets. This measure, aligned with the theoretical interpretation of resilience within the attrition game framework, is utilized as a key indicator for determining the level of strategic effort in the intra-coalition competition.

To visualize the empirical results derived from the proposed compound attrition game framework, two types of outputs were generated based on the daily closing prices of the six selected assets (three cryptocurrencies and three traditional market assets) during the period from January 1, 2018 to December 31, 2023. These outputs serve to illustrate the volatility-adjusted strategic resilience of each asset, their relative rankings within their respective coalitions, and the temporal dynamics of their market behavior. The outputs include:

(1) a tabular summary of the volatility-adjusted Sharpe ratios,

(2) a bar chart ranking the assets by strategic resilience

Volatility-Adjusted Sharpe Ratios (2018–2023)

| | Mean Daily Return | Volatility (Std) | Volatility-Adjusted Sharpe Ratio |
|---|---|---|---|
| Brent_Crude_Oil | 0.0093 | 0.0919 | 0.1007 |
| Solana | 0.0043 | 0.0535 | 0.0803 |
| Bitcoin | 0.0027 | 0.0401 | 0.0685 |
| Ethereum | 0.0013 | 0.0264 | 0.0494 |
| Copper | 0.0007 | 0.0153 | 0.0435 |
| Gold | 0.0002 | 0.0096 | 0.0234 |

**Table 2** – Volatility-Adjusted Sharpe Ratio Table A summary table showing the computed Sharpe ratios adjusted for volatility across six assets (three cryptocurrencies and three traditional market instruments) over the period 2018–2023. This table provides a comparative assessment of strategic resilience across all assets.

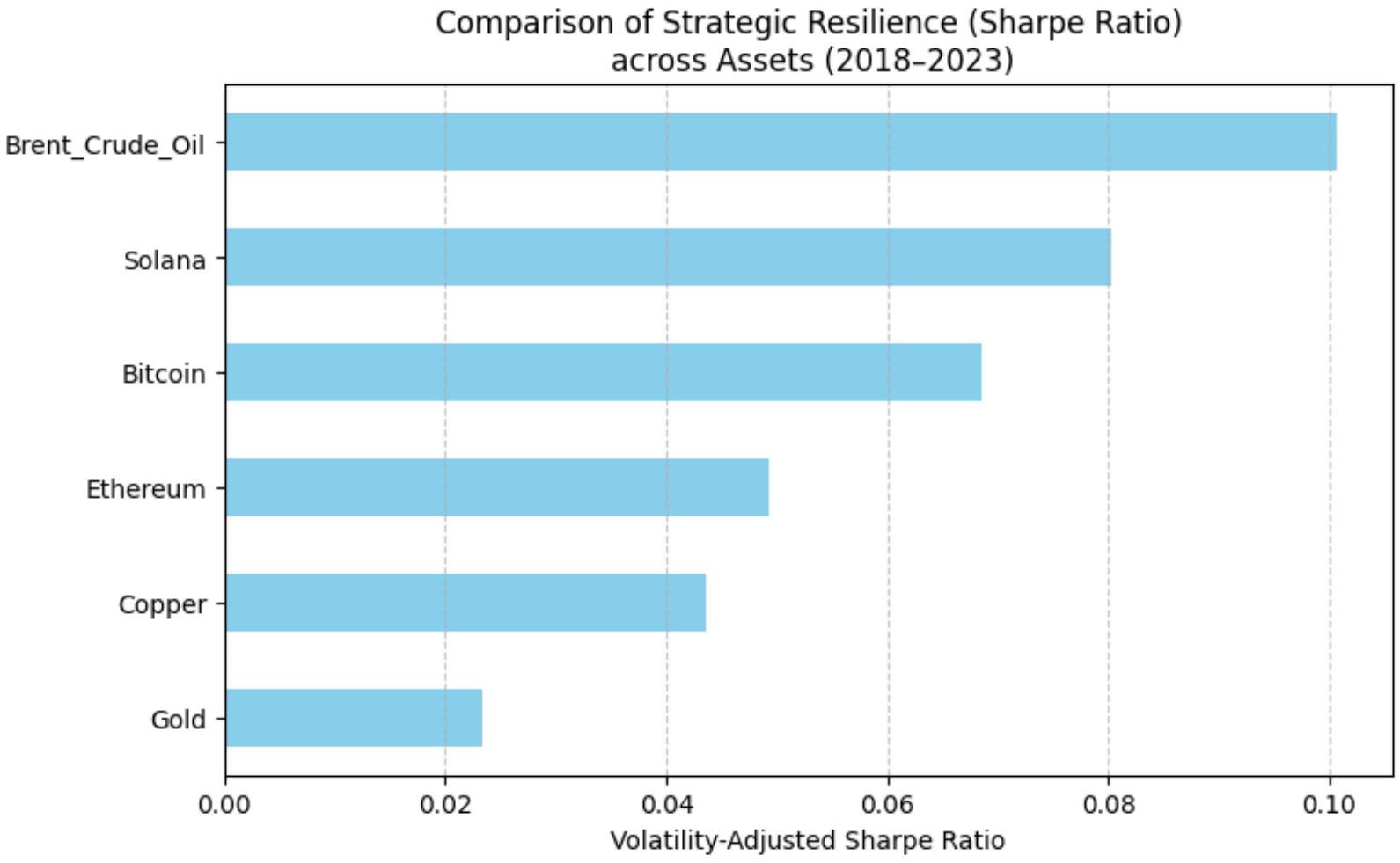

**Figure 1** – Bar Chart of Strategic Resilience A horizontal bar chart illustrating the ranking of all six assets based on their volatility-adjusted Sharpe ratios. This visual aids in identifying which assets demonstrate higher risk-adjusted performance, representing stronger intra-coalition strategic positions.

**Stage One – Coalition Selection (Market Level)**

In this stage, the investor aims to select one of the two primary coalitions representing the markets: the cryptocurrency market or the traditional market. Each coalition comprises a set of assets with varying endurance levels, estimated through financial indicators. To evaluate each coalition, the overall endurance index is calculated as the weighted average of the endurance levels of the coalition's constituent assets. The weights may represent market share or the relative importance of each asset. Subsequently, the investor chooses the coalition with the highest endurance index as their preferred investment option. This decision reflects the inter-coalition competitive game modeled by the proposed compound attrition framework.

To formally connect this decision to the compound attrition model, each market coalition $S_i$ is treated as a strategic agent with a collective endurance capacity. The total endurance index of a coalition is computed as:

$$T(S_i) = \sum_{j \in S_i} w_j . t_j$$

where $t_j$ denotes the endurance level of asset $j$ and $w_j$ reflects its relative importance (e.g., based on market capitalization or liquidity). The index $T(S_i)$ represents the aggregate resilience of the coalition and captures its ability to sustain competitive pressure over time.

In the inter-coalition attrition game, each coalition depletes its resources while competing against the other. Thus, the coalition with the higher $T(S_i)$ is expected to survive longer and dominate the market. The investor, acting rationally under this framework, selects the coalition with the higher endurance index. This choice reflects a first-level strategic move within the compound attrition framework, setting the stage for intra-coalition dynamics in the next phase.

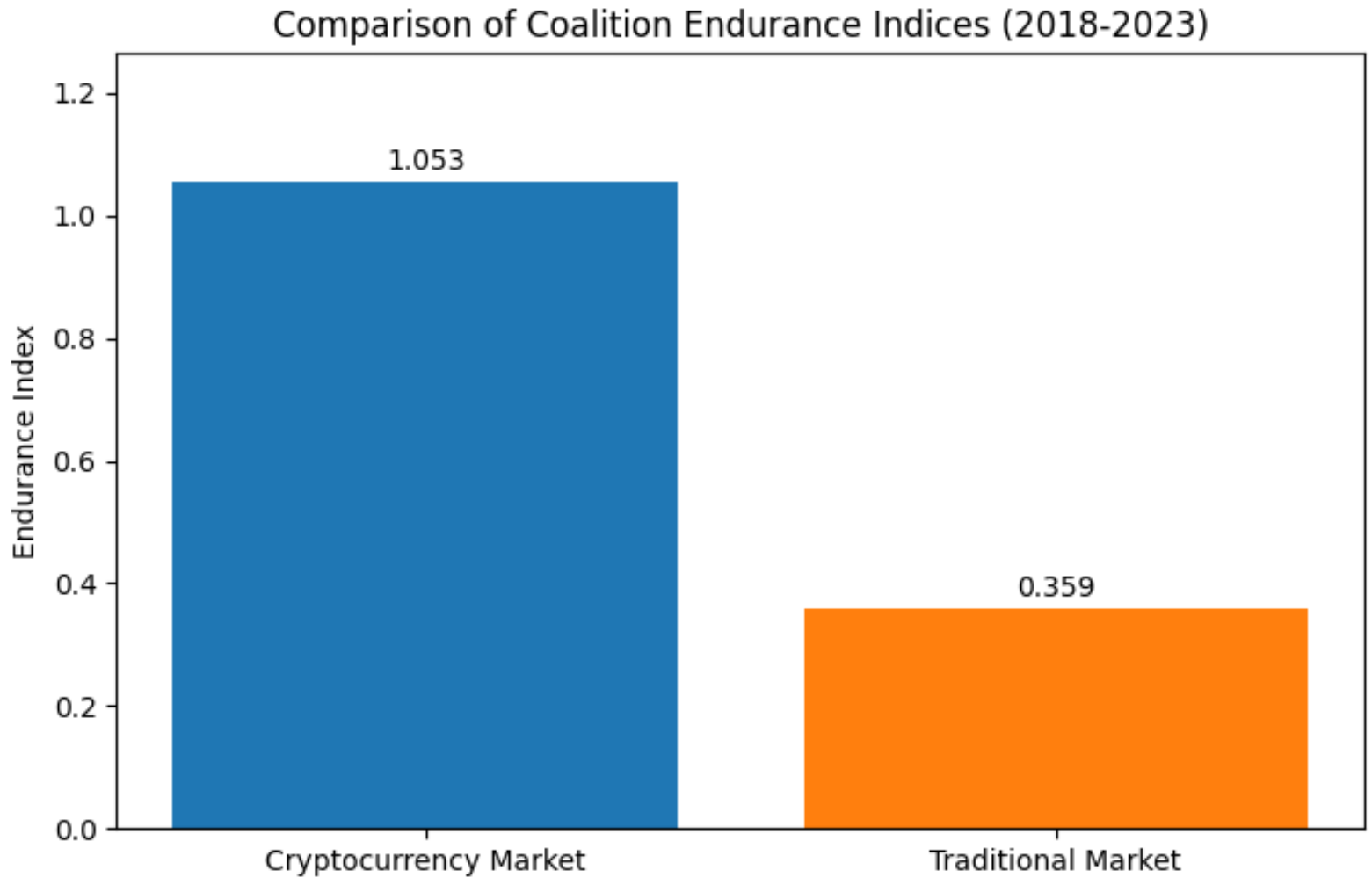

**Figure 2**– Comparison of the overall endurance indices for the cryptocurrency and traditional market coalitions during 2018–2023, illustrating the investor's preference for the market with higher strategic resilience.

Based on the simulation results presented in the figure above, the overall endurance index of the cryptocurrency market during the 2018–2023 period is significantly higher than that of the traditional market. This index, calculated as the weighted average of the endurance levels of each coalition's assets, reflects the relative stability and superior capacity of the cryptocurrency market to withstand competitive and high-risk conditions. Accordingly, in line with the inter-coalition competition model, the investor strategically selects the cryptocurrency market coalition as the preferred investment option.

**Stage Two – Intra-Coalition Competition (Asset Level)**

At this stage, the focus of analysis shifts from inter-coalition competition to intra-coalition rivalry, specifically within the selected coalition—the cryptocurrency market. Unlike the first stage, which examined the strategic dynamics between traditional and crypto markets, this phase explores the competition among individual assets within a single market. The theoretical foundation of this stage is built upon the Attrition Game framework, in which each asset seeks to capture a larger share of the investor's capital by adopting a specific strategic posture.

Within the proposed model, the behavior of each asset is characterized by **Equation (2)**, which defines the utility function ($\pi_i$) of asset $i$. The equation can be interpreted as follows:

$t_i$: Denotes the resilience level of asset $i$, derived from its internal attributes—such as lower volatility, return stability, or historical performance. This variable reflects the asset's capacity to withstand market pressures and sustain itself in the face of competition.

$a_i$ : Represents the attractiveness of asset $i$, determined by external factors such as its expected return, market adoption rate, or investor demand indices.

$S_k$: Refers to the set of assets within coalition $k$ (here, the cryptocurrency market), comprising those that engage in competition during the second stage.

$\pi_i$: Indicates the final share of investment allocated to asset $i$, proportionate to its ability to combine resilience and attractiveness relative to its competitors.

**Equation (2)** analytically illustrates how each asset can attract a larger portion of the total investment pool based on the ratio of its resilience and attractiveness to those of its peers. This intra-coalition competition constitutes the second stage of our composite game model, enabling the investor not only to select the optimal market but also to make an informed and strategic choice among the assets within that market.

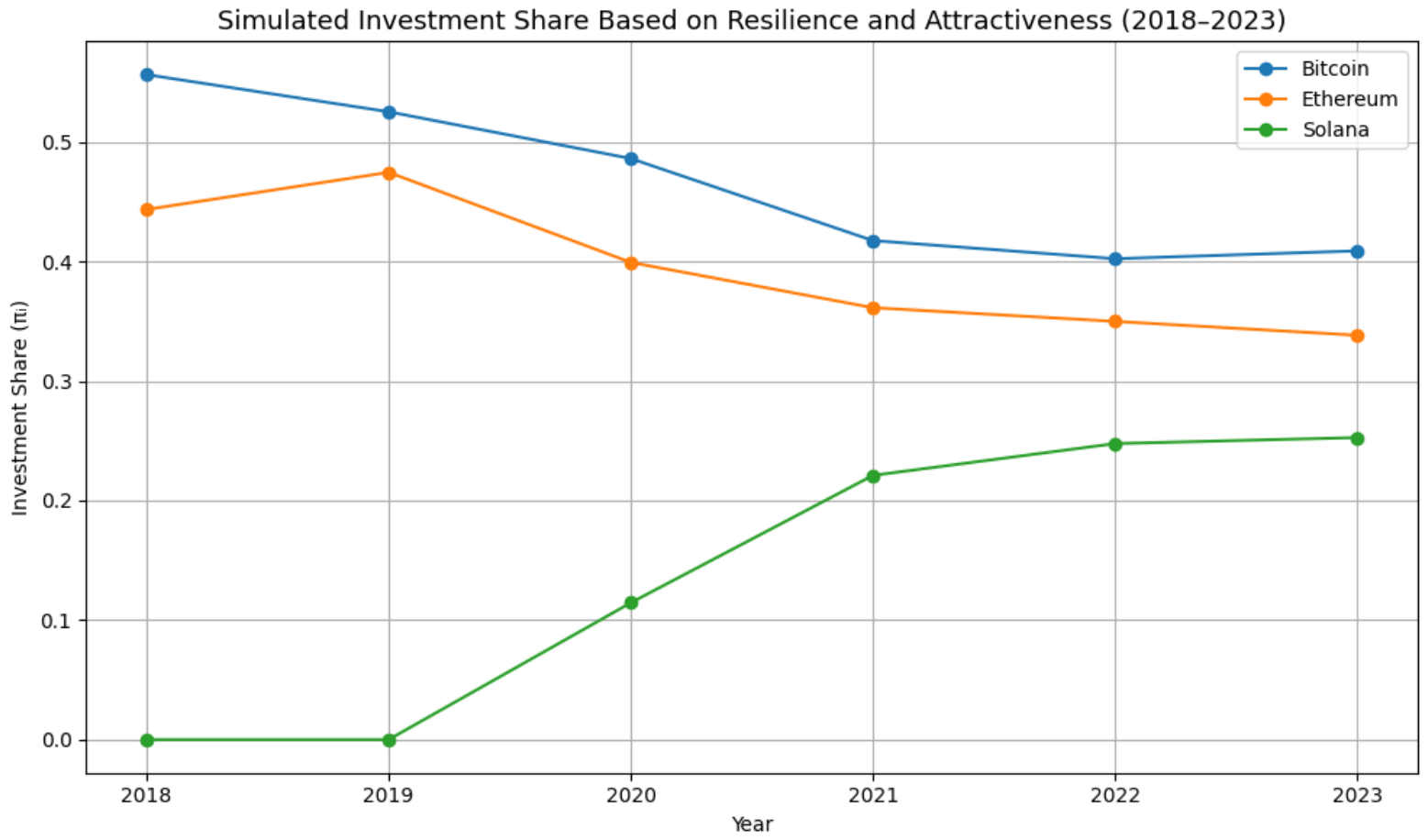

**Figure3**– Simulated investment shares ($\pi_i$) for Bitcoin, Ethereum, and Solana over the period 2018–2023. The allocation is computed based on each asset's relative resilience ($t_i$) and attractiveness ($a_i$), using Equation (2.3). The dynamics reflect how shifts in risk-adjusted appeal and market maturity influence intra-coalition competition among cryptocurrencies.

**Simulation Results and Asset-Level Dynamics (2018–2023)**

The simulation results reveal that the allocation of investment among major cryptocurrencies—Bitcoin, Ethereum, and Solana—over the period 2018 to 2023 has been primarily governed by the interplay of two key factors: each asset's relative resilience ($t_i$) and external attractiveness ($a_i$).

**Early Years (2018–2019):**

During this initial phase, the market consisted solely of Bitcoin and Ethereum, with Solana yet to enter. Owing to its superior combination of resilience and attractiveness, Bitcoin commanded the largest share of investment. Ethereum followed as a close second, albeit still trailing behind Bitcoin in overall investor allocation.

**Year 2020 – Solana's Entry:**

Solana entered the market as a new contender. Although initially exhibiting lower levels of resilience and attractiveness relative to its established rivals, its position gradually strengthened due to rapid technological innovation and growing adoption within DeFi and NFT ecosystems.

**Middle Years (2021–2022):**

During this period, Solana experienced notable improvements in both resilience and attractiveness metrics, enabling it to capture a larger portion of investor capital. While Bitcoin maintained its leading position, the gap between its market share and that of its competitors began to narrow.

**Year 2023:**

By the final year of analysis, the three cryptocurrencies approached a state of relative equilibrium. Bitcoin and Ethereum continued to attract a larger share of investment compared to Solana; however, the overall trend points to intensifying competition and a convergence in investment shares. This progression underscores a broader insight: within a single coalition (in this case, the cryptocurrency market), newer entrants can erode incumbent dominance by systematically improving their performance indicators, thereby attracting an increasing portion of capital over time.

This intra-coalition analysis demonstrates that even after selecting the cryptocurrency market as the dominant coalition in the first stage, an investor must still strategically differentiate among individual assets based on their distinct attributes. The attrition game framework effectively captures the temporal dynamics of this competition, illustrating that final investment shares are a function of relative strengths, rather than absolute superiority.

## Discussion

The previous chapters have introduced the **Compound Coalition-Attrition Game (CCAG)** as a robust theoretical framework and demonstrated its effectiveness in explaining and predicting outcomes in high-risk competitive environments. The primary utility of this model lies in integrating inter-coalition and intra-coalition attrition wars within a single, interconnected structure that reveals critical feedback loops determining the fate of coalitions. While the model's applications in financial market selection and geopolitical competitions serve as compelling validations, the true power of the CCAG extends far beyond these examples. This discussion explores the broader implications of the model, highlights its potential for diverse applications, addresses its limitations, and suggests promising directions for future research. The model is not merely a tool for analyzing past events but offers a versatile lens for understanding the fundamental dynamics of multilayered competition wherever it occurs.

One highly suitable area for future application of the CCAG model lies in the domain of political economy and public policy, particularly in the formation and stability of ruling coalitions. Governments are rarely unified entities; rather, they consist of coalitions of political parties, each with distinct ideologies, electoral bases, and policy priorities. The governance process can be viewed as a continuous composite coalition-attrition game. The inter-coalition prize V represents the ability to implement policy agendas and maintain power. The collective endurance $T_k$ of a ruling coalition reflects its capacity to preserve unity and withstand political pressures, which can be modeled as a coordination function:

$$T_k \; = f\left(t_k \;\right) = \left(\sum a_i t_k^i \;\right) exp(-\gamma \cdot Var\left(t_k^i\right)) \;\; t_i$$

Here, $t_k^i$ denotes the political capital and commitment of a minor coalition partner to a specific policy. The variance term is critical; large disagreements over key issues (such as budget deficits or foreign policy) increase $Var\left(t_k^i\right)$, thereby reducing the overall coalition endurance $T_k$ and making it vulnerable to collapse. The breakdown of ruling coalitions in parliamentary democracies—often triggered by the defection of a minor partner over a contentious issue—is a classic example of a player with low $t_i$ precipitating systemic failure. This model can be employed to simulate the stability of various coalition governments, predict their survival probabilities based on policy platforms, and elucidate the conditions under which larger coalitions prove more stable than minimal winning ones.

Another promising domain for application is ecology and evolutionary biology, particularly in the study of mutualistic and symbiotic relationships. For example, consider the symbiosis between clownfish and sea anemones. This relationship can be conceptualized as a two-player coalition competing against external environmental factors (harsh climate, predators, etc.). The prize V represents survival and reproductive success. The individual endurance $t_i$ for each species corresponds to the energy invested in the symbiosis—protection in the case of the clownfish, and nutrients derived from the clownfish's waste for the sea anemone. The cost $c_i(t_i)$ represents the metabolic expense of providing this benefit. The collective endurance $T_k$ reflects the fitness of the symbiotic pair.

If one species cheats (i.e., reduces its $t_i$) while the other maintains a high level of investment, the partnership becomes unstable, reflecting a breakdown of cooperation. This application can provide a game-theoretic foundation for understanding the evolution and stability of symbiosis, going beyond purely biological explanations by incorporating strategic decision-making—even among non-intelligent actors operating on evolutionary timescales.

This model also holds significant potential in the fields of operations research and supply chain management. Modern supply chains are not linear sequences but complex networks of firms, each with distinct interests. A supply chain can be modeled as a coalition $S_k$ competing against other supply chains (coalitions) for market share. The prize V represents a contract with a major retailer or consumer demand. The collective endurance $T_k$ of the supply chain is its ability to reliably and efficiently meet demand. This is a classic "weakest link" scenario fully captured by the matching function

$$T_k = f(t_k) = \min_{i \in S_k} t_k^i$$

where $t_k^i$ denotes the reliability or capacity of a supplier within the chain. An unreliable supplier (with low $t_k^i$) can incapacitate the entire chain, reducing $T_k$ to zero and causing the chain to lose the contract to a competitor. This provides a formal framework for analyzing supply chain resilience. It can be used to quantify redundancy value (e.g., having multiple suppliers), assess risks of over-dependence on a single source, and determine optimal investments to improve the reliability of the weakest link. Recent global disruptions, from pandemics to geopolitical conflicts, have underscored supply chain fragility, and the CCAG model offers a rigorous approach for analyzing and designing more robust systems.

In the domain of innovation and intellectual property, particularly in the development of complex technology ecosystems, the CCAG model offers a novel perspective. The development of a new technology standard, such as 5G or a new operating system, often involves a consortium of firms. This consortium can be conceptualized as a coalition $S_k$ competing against rival consortia. The prize V represents market adoption and licensing revenues derived from the standard. The individual endurance $t_k^i$ of a member firm corresponds to its investment in research and development and its commitment to the standard. Intra-coalition resources $W_k$ represent the licensing costs. A consortium whose members are fully committed (with high $t_i$) and share a common vision (low $Var(t_i)$) will exhibit high collective endurance $T_k$ and is more likely to see its standard adopted. Conversely, a consortium facing internal conflicts, where members covertly develop incompatible proprietary technologies (high $Var(t_i)$), will exhibit low $T_k$ and is likely to be outcompeted by a more cohesive rival. This framework can assist policymakers and firms in understanding the necessary conditions for successful standard-setting and in designing more effective collaboration agreements.

Despite its wide applicability, the CCAG model possesses inherent limitations that warrant careful consideration. The primary limitation lies in the assumptions of common knowledge and full rationality. In its current form, the model assumes that all players are perfectly rational utility maximizers with complete knowledge of the game structure and the payoffs of other players. In reality, information is often incomplete and asymmetric, and players may exhibit bounded rationality or be motivated by factors beyond simple payoff maximization, such as ideology, fairness, or spite. Future research could extend this model to Bayesian games, where players possess incomplete information regarding the costs or endurance levels of others. This extension is particularly relevant for geopolitical applications, where information is a critical and uncertain factor. Another limitation concerns the static nature of the prize V.

In many real-world scenarios, the value of the prize itself changes over time or depends on the actions of the players. For instance, a protracted standards war can diminish the total available market V , thereby reducing the value of victory. Extending the model to a differential game framework, where V is a function of time and players' strategies, would represent a significant advancement.

In conclusion, the Composite Coalition-Attrition Game model provides a powerful and versatile language for describing and analyzing a broad spectrum of complex competitive interactions. By formally linking the fate of a group to the internal struggles of its members, this model offers a more nuanced and realistic depiction of strategic competition compared to traditional frameworks. The applications discussed here—from political coalitions and biological symbioses to supply chains and technology standards—demonstrate the model's vast potential. While challenges remain in incorporating more realistic assumptions about information and dynamics, the core framework establishes a solid foundation. It is our hope that this model will stimulate further research across disciplines and encourage scholars to explore the intricate interplay of cooperation and conflict that defines many human endeavors—from boardrooms to battlefields, and from ecosystems to the global stage. The true test of any theory lies not only in its sophistication but in its ability to illuminate the world, and the CCAG model casts a novel and illuminating light on the complex attrition wars shaping our world.

## Conclusion

In this paper, a novel model entitled the Compound Coalition-Attrition Game (CCAG) was introduced, which simultaneously and integratively models both inter-coalition and intra-coalition competition within the framework of war of attrition logic. This two-layer dynamic model fills a significant gap in the theoretical literature, where internal and external competitions have traditionally been analyzed separately and in a simplified manner. Theoretical analyses demonstrated that strategic equilibria at both layers interact and influence each other, and this interplay can generate reinforcing cycles of success or failure.

Simulations based on real-world data from traditional markets (gold, oil, copper) and the cryptocurrency market (Bitcoin, Ethereum, Solana**)** further showed that the CCAG model can accurately replicate the competition trajectory and predict the conditions under which the leading player changes. Moreover, counterfactual analyses revealed the impact of variations in resources and costs on competition outcomes. These findings not only validate the model but also demonstrate its practical applicability in analyzing complex, multi-layered competitions in economic, financial, and technological domains.
The most significant innovation of this model lies in the integration of cooperative game theory and war of attrition logic, allowing for a realistic representation of both inter-coalition competition and the complex, attritional dynamics of intra-coalition rivalry. This approach enables more precise analysis and better prediction of player behaviors in environments where cooperation and competition coexist with mutual influence.
Given the increasing importance of emerging markets and the complexity of strategic interactions within them, the CCAG model can serve as an effective tool for policymakers, investors, and researchers in risk analysis, coalition management, and optimal decision-making. Furthermore, future extensions of the model incorporating timing factors, incomplete information, and multiple players could enhance its depth and applicability.
Ultimately, this study takes a meaningful step toward bridging the existing theoretical and practical gaps and lays the groundwork for subsequent research in cooperative games and wars of attrition with a multi-layered perspective.

**Author Contributions:**
Professor Madjid Eshaghi Gordji supervised the project and provided valuable guidance that helped develop and refine the research idea.
Mohamadali Berahman was responsible for writing the manuscript and compiling the materials.
**Competing interests**
The author(s) declare no competing interests